\documentclass[final,3p,times,twocolumn, sort&compress]{elsarticle}
\usepackage{layouts}
\usepackage{amsmath,amssymb,stmaryrd,amstext,amsthm}
\usepackage{mathtools}
\usepackage{bm}	
\usepackage{caption}
\usepackage{siunitx}
\usepackage{booktabs}
\usepackage{multirow, graphicx}
\usepackage[T1]{fontenc}

\usepackage{lscape}
\usepackage{pdflscape}
\usepackage[table, svgnames, dvipsnames]{xcolor}
\usepackage{longtable}
\usepackage{tabulary}
\usepackage{rotating,array,ragged2e,caption}
\usepackage{tabularx}
\usepackage{subcaption, nicefrac, lineno, float, siunitx}
\usepackage{enumitem}
\usepackage{adjustbox}

\usepackage{array}
\usepackage{makecell, cellspace, caption}
\usepackage{hyperref}
\hypersetup{
    colorlinks=true,
    linkcolor=blue,
    filecolor=magenta,      
    urlcolor=cyan,
    pdftitle={Global improvement and performance evaluation of two-dimensional progressive data-augmented turbulence models},
    }

\usepackage[english]{cleveref}

\newcommand{\blue}{\color{black}}
\newcommand{\black}{\color{black}}

\newcommand{\volsym}{\rlap{\kern.08em--}V} % Volume symbol

\def\tsc#1{\csdef{#1}{\textsc{\lowercase{#1}}\xspace}}
\tsc{WGM}
\tsc{QE}
\tsc{EP}
\tsc{PMS}
\tsc{BEC}
\tsc{DE}

\journal{arXiv} 

\begin{document}

\begin{frontmatter}



\title{A generalisable data-augmented turbulence model with progressive and interpretable corrections \blue for incompressible wall-bounded flows\black}

\author[inst1,inst2]{Mario Javier Rinc\'on\corref{cor1}}
\cortext[cor1]{Corresponding authors}
\ead{mjrp@mpe.au.dk}
\author[inst2]{Martino Reclari}
\author[inst3]{Xiang I. A. Yang \corref{cor2}}
\author[inst1]{Mahdi Abkar\corref{cor1}}
\ead{abkar@mpe.au.dk}

\address[inst1]{Department of Mechanical and Production Engineering, Aarhus University, 8200 Aarhus N, Denmark}
\address[inst2]{Quality \& Sustainability Department, Kamstrup A/S, 8660 Skanderborg, Denmark}
\address[inst3]{Department of Mechanical Engineering, Pennsylvania State University, State College, PA, 16802, USA}

\begin{abstract}
\noindent The integration of interpretability and generalisability in data-driven turbulence modelling remains a fundamental challenge for computational fluid dynamics applications. 
This study yields a generalisable advancement of the $k$-$\omega$ Shear Stress Transport (SST) model through a progressive data-augmented framework, combining Bayesian optimisation with physics-guided corrections to improve the predictions of anisotropy-induced secondary flows and flow separation simultaneously. Two interpretable modifications are systematically embedded: 1) a non-linear Reynolds stress anisotropy correction to enhance secondary flow predictions, and 2) an activation-based separation correction in the $\omega$-equation, regulated by an optimised power-law function to locally adjust turbulent viscosity under adverse pressure gradients. The model is trained using a multi-case computational fluid dynamics-driven \textit{a posteriori} approach, incorporating periodic hills, duct flow, and channel flow to balance correction efficacy with baseline consistency. Validation across multiple unseen cases -- spanning flat-plate boundary layers, high-Reynolds-number periodic hills, and flow over diverse obstacle configurations -- demonstrates enhanced accuracy in velocity profiles, recirculation zones, streamwise vorticity, and skin friction distributions while retaining the robustness of the original $k$-$\omega$ SST in attached flows. Sparsity-enforced regression ensures reduced parametric complexity, preserving computational efficiency and physical transparency. Results underscore the framework's ability to generalise across geometries and Reynolds numbers without destabilising corrections, offering a validated framework toward deployable, data-augmented turbulence models for numerical simulations.
\end{abstract}

% \begin{highlights}
% \item A novel progressive data-augmented turbulence model has been developed following \textit{a posteriori} training.
% \item The model displays interpretable corrections, is numerically stable, generalisable, and accurate.
% \item The new model shows superior interpretability and generalisability performance compared with state-of-the-art machine learning models.

% \end{highlights}

\begin{keyword}
Turbulence modelling \sep Computational fluid dynamics \sep RANS \sep Data-driven \sep Generalisability %
\end{keyword}

\end{frontmatter}

% \linenumbers

\section{Introduction}

The development of turbulence models that can accurately predict complex fluid behaviour while maintaining model consistency, interpretability, and generalisability remains a fundamental challenge in fluid mechanics \cite{DURAISAMY2025311, brunton2020machine}. Among the Reynolds-Averaged Navier-Stokes (RANS) turbulence models, linear eddy-viscosity models, such as the Spalart-Allmaras \cite{spalart1992one} and the $k-\omega$ Shear Stress Transport (SST) models \cite{menter2003ten}, are widely used due to their robustness, computational efficiency, and interpretability. However, these models often fail to capture certain turbulent flow intricacies, particularly in complex geometries and highly unsteady flows \cite{bin2024large,raiesi2011evaluation}.

Compared to high-fidelity methods such as direct numerical simulations (DNS) and large-eddy simulations (LES), RANS remains the preferred approach for industrial computational fluid dynamics (CFD) applications due to its cost-effectiveness \cite{menter2021overview, corson2009industrial, li2022grid}. The accuracy of RANS simulations depends on the performance of turbulence models in predicting the Reynolds stress tensor (RST). However, empirical models such as $k-\varepsilon$ and $k-\omega$ struggle with specific flow phenomena, including secondary flows \cite{nikitin2021prandtl} and boundary layer separation and reattachment \cite{slotnick2014cfd}. 

To address these limitations, machine learning techniques have increasingly been integrated into RANS turbulence modelling, offering the potential to enhance accuracy and efficiency while maintaining physical constraints \cite{duraisamy2021perspectives, liu2021iterative}. However, this integration presents significant challenges, particularly in ensuring interpretability and generalisability \cite{mandler2024generalization}. \blue The turbulence modelling community widely recognises that achieving a fully generalisable data-driven model (one that performs reliably across all flow physics and geometries) remains an extraordinary undertaking and a long-term goal \cite{spalart2015philosophies, spalart2023old, brunton2020machine, brunton2016discovering,chen2023quantifying}. Progress towards this objective often involves methodological advancements that demonstrate robust performance improvements for specific, challenging classes of flows. A key challenge is therefore to develop frameworks that allow for targeted, non-detrimental augmentation for distinct physical phenomena, validated on a range of unseen but related cases. This controlled demonstration of generalisability for targeted improvements, without compromising the baseline model's performance where it is already reliable, is a critical step forward and one that many current data-driven approaches find challenging.\black

Recent advances in data-driven methods have inspired research aimed at refining RANS models by leveraging high-fidelity data \cite{duraisamy2019turbulence}. Most studies focus on \textit{a priori} training to improve RST predictions \cite{tracey2013application, wang2017physics, wu2018physics, ling2016reynolds, kaandorp2020data, mcconkey2022deep, singh2017machine, holland2019field, cruz2019use, weatheritt2016novel, weatheritt2017development, schmelzer2020discovery, amarloo2022frozen, amarloo2023data, hu2025data}, with concerns regarding model generalisability and consistency for unseen cases \cite{sandberg2022machine,bin2024adapting,vadrot2023survey}. These machine learning-based approaches leverage high-fidelity numerical simulations to enhance RANS models. Typically, field inversion techniques or frozen turbulence simulations establish a corrective field, $\beta$, which correlates high- and low-fidelity simulation data. Data-driven methods are then employed to determine model parameters that reproduce the corrective field \cite{singh2017augmentation}. However, these \textit{a priori} training methods pose fundamental challenges related to the ill-conditioning of RANS equations, generalisability, and numerical stability. Since these models are trained solely to match high-fidelity data, they fail to account for feedback effects between the turbulence model and the flow solver, potentially leading to robustness issues during full numerical simulations. Furthermore, their implementation requires careful attention to ensure compliance with the physical constitutive laws governing RANS equations \cite{rumsey2022search}.

Recent studies have sought to address the generalisation problem. Fang \textit{et al.}~\cite{fang2023toward} proposed incorporating a broader set of training cases within the optimisation process to improve model robustness. However, their results indicate that trained models still exhibit limitations when applied to either wall-bounded or free-shear flows. Ho \textit{et al.}~\cite{ho2024probabilistic} introduced a Gaussian Process Emulator for $k$-$\omega$ SST, allowing the model to revert to uncorrected turbulence predictions when operating beyond the training dataset's scope. Similarly, Tang \textit{et al.}~\cite{tang2023data} demonstrated that Bayesian deep learning enhances uncertainty quantification and generalisation through nonlinear corrections. Cherroud \textit{et al.}~\cite{cherroud2025space} presents a machine learning methodology for learning and aggregating customised turbulence models (experts) for selected classes of flows, in order to make predictions of unseen flows with quantified uncertainty. Nevertheless, none of these approaches have successfully achieved full generalisability across diverse flow conditions and geometries. Furthermore, learned corrections are often non-local, which can degrade model performance even in regions where the baseline model is accurate.

A critical limitation of data-driven models is their lack of generalisability beyond training cases. Even for simple canonical cases, these models may fail to reproduce the expected results. Therefore, an alternative approach involves model-consistent training, where turbulence models are optimised based on \textit{a posteriori} performance, ensuring numerical stability and robustness \cite{duraisamy2021perspectives}. Optimisation algorithms, including slope followers \cite{barzilai1988two}, simplex methods \cite{nelder1965simplex}, multi-objective evolutionary algorithms \cite{coello2007evolutionary}, and particle swarm techniques \cite{eberhart1995particle}, have been applied to CFD-driven optimisation with promising results \cite{zhao2020rans, saidi2022cfd}. However, computational expense remains a significant challenge. To mitigate this, surrogate-based optimisation (SBO) techniques have been introduced, using response surfaces to approximate objective functions efficiently \cite{myers2016response, queipo2005surrogate}. Despite their advantages, SBO methods have not been extensively applied to RANS turbulence model improvement, with only a few studies exploring their potential \cite{zhao2020rans, waschkowski2022multi, saidi2022cfd}.

Regarding \textit{a posterior} training, Bin \textit{et al.}~\cite{bin2022progressive, bin2023data, bin2024constrained} proposed a progressive augmentation approach, which systematically incorporates corrections without compromising baseline model performance. Recent studies by Raje \textit{et al.} \cite{raje2024recent} and Oulghelou \textit{et al.} \cite{oulghelou2024machine} have likewise prioritised the implementation of blending, flow clustering, and diverse data-driven techniques to reach similar results in two-dimensional cases, although no general, interpretable model or methodology consensus is achieved to \blue democratise \black the general use of the findings.

\begin{figure*}[t!]
    \centering
    \includegraphics[width=1\linewidth]{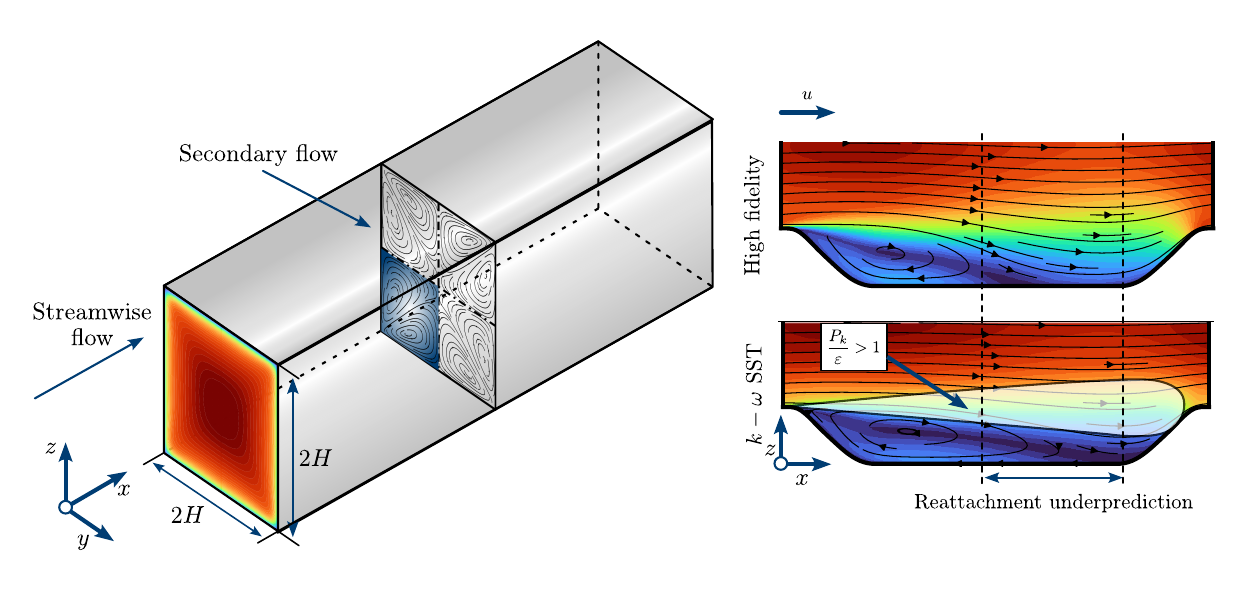}
    \caption{High-fidelity results of anisotropy-based secondary flow in duct flow case (left), and streamwise velocity and stream function of flow separation at periodic hills case (right) with RANS $k$-$\omega$ SST turbulence model and high-fidelity simulation results. The ratio of the turbulent kinetic energy production to dissipation $\frac{P_{k}}{\varepsilon}$ is used as an indication of flow separation.}
    \label{fig:Fig1}
\end{figure*}

To enhance interpretability, data compression techniques such as principal component analysis have been employed to reduce dataset dimensionality and extract dominant flow features, striving for a unique features-to-augmentation map. By integrating such techniques with machine learning, models can achieve improved prediction accuracy while preserving interpretability. Specifically, progressive data-augmented (PDA) turbulence models \cite{rincon2023progressive, amarloo2023progressive}, implemented within the OpenFOAM Turbulence Community database \cite{weller1998tensorial}, have demonstrated robustness in predicting anisotropy-induced secondary flows and flow separation in two-dimensional geometries without degrading performance in canonical flows. However, their validation has been limited to specific cases, including Prandtl's secondary flow of the second kind and boundary layer separation over bumps, hills, and convergent-divergent ducts. The necessity for broader evaluation across diverse flow conditions has been emphasised in recent studies \cite{cinnella2024datadriven, li2024enhancing}, highlighting the need for a systematic improvement and performance assessment of PDA models. Moreover, although PDA models have demonstrated accurate predictions of these two flow phenomena individually, it is not trivial that they can be effectively combined into a single model capable of simultaneously predicting secondary flows and separation with consistent accuracy.

Hence, to address the challenges of model consistency, interpretability and generalisability in turbulence modelling, this study employs a progressive augmentation approach to enhance the standard $k$-$\omega$ SST model with data-driven correction factors, specifically targeting anisotropy-induced secondary flows and flow separation phenomena (Fig. \ref{fig:Fig1}). The methodology, detailed in Section \ref{sec:methodology}, introduces a correction framework comprising a non-linear Reynolds stress anisotropy modification embedded within the velocity transport equation and an activation-based correction in the $\omega$-equation to refine turbulence behaviour under adverse pressure gradients. Model training is conducted using an \textit{a posteriori} multi-case CFD-driven optimisation framework, incorporating duct flow, periodic hills, and channel flow cases to ensure a balanced correction strategy. Section \ref{sec:results} presents the verification and validation of the trained model on unseen test cases, encompassing a range of Reynolds numbers and geometric configurations. Finally, Section \ref{sec:conclusions} summarises the findings and discusses the implications of the proposed framework for data-driven turbulence modelling. \blue The proposed PDA framework is developed following established guidelines for data-driven turbulence modelling, such as those defined by Spalart \cite{spalart2015philosophies, spalart2023old}. The new model aims to leverage the strengths of a well-validated baseline RANS model ($k-\omega$ SST) by introducing physics-informed, interpretable corrections with a limited number of optimisable parameters. The \textit{a posteriori} training methodology ensures numerical stability and solver-consistency, while the design of the correction terms, particularly an activation function for the separation correction, aims to preserve the baseline model's accuracy in well-predicted flow regimes and ensure robust behaviour. This approach consciously tries to avoid pitfalls associated with purely data-driven black-box models and prioritises generalisability and physical consistency. \black By systematically ensuring model consistency, interpretability, and generalisability, this work advances the integration of physics-informed data augmentation within RANS turbulence modelling, ensuring reliability across a broad range of flow conditions while preserving physical interpretability.

\blue It is important to clarify the scope of the present investigation. The developed framework and the resulting model are specifically focused on improving predictions for incompressible, internal and wall-bounded flows. The data-driven corrections are tailored to address deficiencies observed within this flow class. For instance, the under-estimation of turbulent viscosity in the separated internal flow cases studied here is a known issue that the proposed correction targets. This behaviour is notably distinct from that observed in many external aerodynamic applications, where standard RANS models tend to over-predict turbulent viscosity, leading to delayed stall predictions. Consequently, the direct application of the present model is confined to the class of flows for which it was developed. The extension of this progressive augmentation methodology to other regimes, such as external aerodynamics or compressible flows, represents a valuable direction for future work but is beyond the scope of this study.\black

\section{Methodology}
\label{sec:methodology}

This section outlines the methodology for progressively augmenting the RANS model to correct anisotropy in the RST and improve flow separation predictions using CFD-driven optimisation techniques. The section is developing upon the methods used in \cite{rincon2023progressive, amarloo2023progressive} (where the reader is referred for further details), while elaborating in the novel approaches included in this study. 

In the baseline model used, the RANS equations for an incompressible steady flow employing Reynolds decomposition for velocity and pressure, are given as \cite{pope2000turbulent}:

\begin{equation}
\label{eq:ransDC}
    u_i = \langle u_i \rangle + u^{\prime}_i, \hspace{0.4cm}  p = \langle p \rangle + p^{\prime},   
\end{equation}

\begin{equation}
\label{eq:ransEq}
\partial_i \langle u_i \rangle = 0, \hspace{0.4cm} \partial_j \left( \langle u_i \rangle \langle u_j \rangle \right) = -\frac{1}{\rho}\partial_i \langle P \rangle + \partial_j \left(\nu \partial_j \langle u_i \rangle - A_{ij}\right),\end{equation}

\noindent where $i, j = 1, 2, 3$ indicate streamwise ($x$), spanwise ($y$), and vertical ($z$) directions. The variables $u_i$ and $p$ represent velocity components and kinematic pressure, respectively, decomposed into mean values $\langle \cdot \rangle$ and fluctuations $\cdot^{\prime}$. The modified kinematic pressure is defined as $\langle P \rangle = \langle p \rangle + \frac{1}{3}\rho\langle u^{\prime}_{i} u^{\prime}_{i} \rangle$, and the anisotropic part of the RST is expressed as:

\begin{equation}
    A_{ij} =\langle u^{\prime}_{i}u^{\prime}_{j} \rangle - \frac{1}{3}\langle u^{\prime}_{k} u^{\prime}_{k} \rangle \delta_{ij}.
\end{equation}

Applying the Boussinesq approximation, $A_{ij}$ is modelled as

\begin{equation}
    A_{ij} = -2\nu_t S_{ij},
    \label{eq:eddyViscosityModel}
\end{equation}

\noindent where $S_{ij} = \frac{1}{2} \left( \partial_i\langle u_j \rangle + \partial_j \langle u_i \rangle \right)$ is the mean strain-rate tensor, and $\nu_t$ is the turbulent viscosity given by the $k$-$\omega$ SST model \cite{menter1994two} as

\begin{equation}
    \nu_t = \frac{a_1 k}{\text{max} \left(a_1 \omega, F_2 S \right)}.
    \label{eq:nutModel}
\end{equation}
Here, $k$ is the turbulent kinetic energy, and $\omega$ is the specific dissipation rate. The parameter $a_1$ is a model constant, with a value of 0.31, and $F_2$ is a blending function defined in the original $k$-$\omega$ SST model \cite{menter1994two}.

\subsection{Reynolds stress anisotropy correction}
To improve secondary flow predictions, the linear eddy-viscosity model (Eq.~\ref{eq:eddyViscosityModel}) is extended using Pope's decomposition \cite{pope1975more} and adding the non-linear normalised second basis tensor
$T_{ij}^{(2)}=\left({S_{ik}\Omega_{kj}-\Omega_{ik}S_{kj}}\right)/{\omega^2}$ 
with an unknown correction function $\alpha_{A}$ as follows:

\begin{equation}
    A_{ij} = -2\nu_{t}\left( S_{ij} - \alpha_{A} \frac{S_{ik}\Omega_{kj}-\Omega_{ik}S_{kj}}{\omega} \right),
    \label{eq:rstModel}
\end{equation}

\noindent where $\Omega_{ij} = \frac{1}{2}\left(\partial_i\langle u_j \rangle - \partial_j \langle u_i \rangle\right)$ is the mean rotation-rate tensor. \blue Specifically, this modified $A_{ij}$ is incorporated into the mean momentum transport equations (Eq. \ref{eq:eddyViscosityModel}). Subsequently, the turbulent production term, $P_k = A_{ij}S_{ij}$, is calculated using this modified $A_{ij}$. Finally, this production term $P_k$, now influenced by the $\alpha_A$ correction, is then used as the production term in both the $k$-transport equation and the $\omega$-transport equation. \black The addition of these two variables ($\alpha_{A}$ and $T_{ij}^{(2)}$) ensures the preservation of the canonical model's turbulent viscosity prediction while only enhancing Reynolds stress anisotropy, and therefore, Prandtl's second type of secondary flow. \blue While a full tensor basis expansion for 2D flows includes an additional term to fully represent the Reynolds-stress anisotropy \cite{pope1975more, gatski1993explicit}, this study focuses only on the integration of the second tensor basis due to its established role in predicting anisotropy-induced secondary flows \cite{rincon2023progressive, amarloo2023progressive}, and to manage computational optimisation costs. Future work could explore the inclusion of further bases. \black

Finally, the unknown correction function $\alpha_{A}$ is determined using \textit{a posteriori} CFD-driven optimisation.

\subsection{Flow separation correction}
To improve flow separation predictions while preserving the model's self-similar behaviour, the transport equation for $\omega$ is carefully modified by introducing an additional correction term, $R$ \cite{amarloo2023progressive}, following a physics-based reasoning grounded in the mathematical structure of the equation:

\begin{equation}
    \partial_j \left(\langle u_j \rangle \omega \right) = \frac{\gamma}{\nu_t}\left(P_k + R\right) - \beta \omega^2  + \partial_j \left(\left(\nu+\sigma_{\omega}\nu_t\right) \partial_j \omega \right) +  CD_{k\omega}.
    \label{eq:newOmegaModel}
\end{equation}

The correction term $R$, defined in \cite{amarloo2023progressive}, is defined as:

\begin{equation}
\label{eq:correctionModel}
R = \partial_i\langle u_j \rangle \left(2\nu_t \alpha_{S} S_{ij}\right) \mathcal{X} = P_k \alpha_{S} \mathcal{X},
\end{equation}

\noindent where $P_{k}$ is the turbulent kinetic energy production, and $\alpha_{S}$ is an unknown separation correction function, determined again through a CFD-driven optimisation process. Here, $\mathcal{X}$ is an activation function based on shear stress transport (SST) reflection inside the $k-\omega \text{SST}$ model.

\begin{figure}[!t]
    \centering
    \includegraphics[width=1\linewidth]{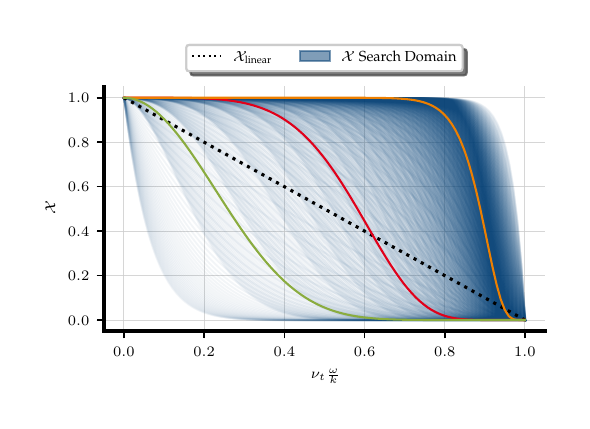}
    \caption{Search area for $\mathcal{X}$. Blue translucent lines represent possible activation functions with diverse combinations of $\lambda_{1}$ and $\lambda_{2}$ in the search domain.\blue The linear function $\mathcal{X}_\text{linear}$ shows the case where $\lambda_{1} = \lambda_{2} = 1$\black. Three examples of possible activation functions are given. Red: $\lambda_{1} = 5$, $\lambda_{2} = 10$. Orange: $\lambda_{1} = 20$, $\lambda_{2} = 7$. Green: $\lambda_{1} = 2$, $\lambda_{2} = 10$.}
    \label{fig:Fig2}
\end{figure}

\blue This approach is not a conventional machine-learning design but is derived from a detailed analysis of the governing equations, ensuring consistency with underlying physical principles. The correction term directly targets regions where linear eddy-viscosity models struggle, particularly under adverse pressure gradients where equilibrium assumptions fail \cite{menter1994two}. To ensure the correction is both effective and selective, the activation function $\mathcal{X}$ is designed to act as a switch, engaging the correction term only where physically necessary. The design is based on the behaviour of the baseline $k-\omega$ SST model itself, particularly the relationship between turbulent viscosity $\nu_t$, turbulent kinetic energy $k$, and specific dissipation rate $\omega$. As observed by Menter \cite{menter2003ten}, in well-behaved equilibrium boundary layers, $\nu_t \approx k/\omega$. Under these conditions, the correction should be inactive. Conversely, in regions of strong adverse pressure gradients and flow separation, the SST model's limiters often cause $\nu_t$ to be significantly smaller than $k/\omega$. To leverage this behaviour, a power-law activation function is defined as follows:

\begin{equation}
    \mathcal{X} = \left(1 - \left( \nu_t \frac{\omega}{k}\right)^{\lambda_1} \right)^{\lambda_2},
    \label{eq:activationFunctionPower}
\end{equation}

\noindent where $\lambda_1$ and $\lambda_2$ are additional optimisation parameters. This function is monotonically decreasing with respect to the term $\nu_t\omega/k$. When the flow is attached, $\nu_t\omega/k \approx 1$, which drives $\mathcal{X}$ towards zero, effectively deactivating the correction $R$ and recovering the standard model's behaviour. However, in separated regions where $\nu_t\omega/k < 1$, $\mathcal{X}$ becomes positive, switching on the correction. The exponents $\lambda_1$ and $\lambda_2$ are optimised to control the sharpness and strength of this transition, ensuring a numerically stable activation. This physics-guided design preserves the model's predictive consistency in attached flows while targeting its known deficiencies in separated flows. This behaviour can be seen in more detail in Fig. \ref{fig:Fig2}.\black

% \noindent where $\lambda_1$ and $\lambda_2$ are additional optimisation parameters to ensure the accurate implementation of the separation correction (Fig. \ref{fig:Fig2}). This function regulates the activation of the correction term, ensuring it only engages where local corrections are necessary, thus preserving the model's predictive consistency in attached flows.

By grounding the modification in a more conventional mathematical reasoning and physical insight, this approach ensures that the correction term enhances predictive capability without compromising the foundational structure of the $k$-$\omega$ SST model.

\subsection{Sparse regression model}
Following the expanded eddy-viscosity formulation by Pope \cite{pope1975more} and its application to two-dimensional flows \cite{gatski1993explicit, schmelzer2020discovery}, which is the focus of this stuidy, $\alpha_{A}$ and $\alpha_{S}$ are defined by the first two normalised flow invariants: 

\begin{equation}
    I_{1} = \frac{\text{tr}(S_{ik}S_{kj})}{\omega^{2}},\quad I_{2} = \frac{\text{tr}(\Omega_{ik}\Omega_{kj})}{\omega^{2}},
\end{equation}

\begin{figure*}[t!]
    \centering
    \includegraphics[width=1.0\textwidth]{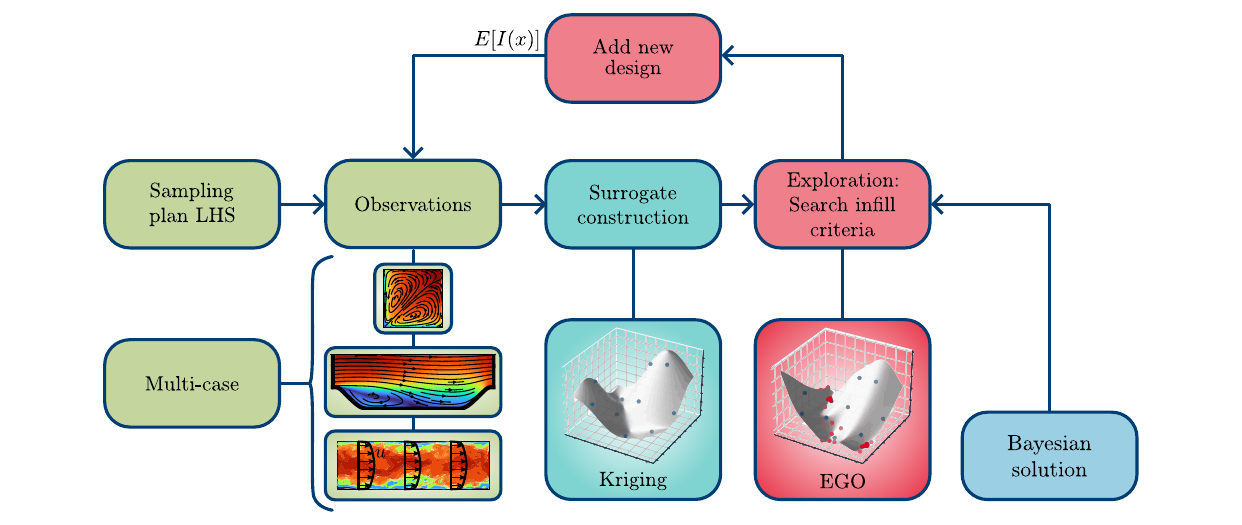}
    \caption{Optimisation approach involving an initial sampling plan using Latin-hypercube sampling, resolved through CFD. A multi-case methodology is adopted, wherein observations are simultaneously conducted across three cases: DF$_{3500, \ AR=1}$, PH$_{2800}$, and CF$_{590}$. Subsequent to this, an initial surrogate model is formulated using Kriging and improved using Bayesian optimisation techniques.}
    \label{fig:Fig3}
\end{figure*}

\noindent where $I_{1}$ and $I_{2}$ are the first and second normalised flow invariants. The final candidate functions used to define the correction functions are standardised with the data of these flow invariants by their z-score to ensure interoperability and promote sparsity, yielding

\begin{equation}
    \mathcal{I}_i=\frac{I_i - \mu_i }{\sigma_i},
\end{equation}

\noindent where $\mathcal{I}_i$ are the standardised candidate functions, and $\mu_{i}$ and $\sigma_{i}$ are the mean and standard deviation of each of the candidate functions $I_{i}$ and cases. The linear combination of these two invariants with unknown coefficients are used following the implementation from \cite{rincon2023progressive, amarloo2023progressive} by a sparse regression method. This simplification yields the final form of the coefficients as design variables for the optimisation as follows:

\begin{subequations}
    \begin{equation}
        \alpha_{A} = A_{0} + A_{1} \mathcal{I}_{1} + A_{2} \mathcal{I}_{2} 
    \end{equation}
    \begin{equation}
        \alpha_{S} = C_{0} + C_{1} \mathcal{I}_{1} + C_{2} \mathcal{I}_{2} 
    \end{equation}
\end{subequations}

\subsection{Bayesian optimisation}

Subsequently, the coefficients $A_i$ and $C_{i}$ are obtained through a CFD-driven optimisation process to improve flow separation and anisotropic stress prediction while preserving the original capabilities of the $k$-$\omega$ SST model. This \textit{a posteriori} optimisation approach includes the use of Kriging surrogate modelling and Bayesian optimisation by leveraging the use of the expected improvement ($E\left[ I(x) \right]$) function \cite{jones2001taxonomy, jones1998efficient, mockus1978application}. The surrogate relies on the principles of design and analysis of computer experiments \cite{sacks1989designs, sobester2008engineering, hastie2009elements}, where diverse applications have been tested against with successful results \cite{rincon2023validating, kawai2014kriging}. 

The optimisation follows the creation of a \textit{space-filling} sampling plan using Latin-hypercube sampling \cite{mckay2000comparison}, which is optimised through the enhanced stochastic evolutionary algorithm \cite{jin2005efficient, damblin2013numerical}. A set of initial samples $n_{0}=50\mathcal{K}$ samples are used to populate the hypercube, where $\mathcal{K}$ are the number of design variables (3 for the RST anisotropy correction, and 5 for the separation correction), subsequently, a number of samples for the Bayesian optimisation $n_{\text{EGO}}=n_{0}$ are generated, leveraging the use of the expected improvement function to reduce surrogate uncertainty and exploring possible global minima locations. A schematic summarising the optimisation methodology is depicted in Fig. \ref{fig:Fig3}.

The optimisation approach follows a \textit{progressive} method of, first, optimise the RST anisotropy correction -- $A_{0}$, $A_{1}$, $A_{2}$ --, and second, optimise the separation correction  -- $C_{0}$, $C_{1}$, $C_{2}$, $\lambda_{1}$, $\lambda_{2}$ -- fixing the RST anisotropy coefficients found in the first optimisation step.

\subsection{Training and validation cases}
In order to train the new model in a holistic manner, well-established two-dimensional canonical cases with diverse flow characteristics are advised to be used. Taking these into consideration, the cases selected for training and validation in this study are summarised in Table~\ref{tab:training_validation_cases}.

\begin{table*}[t!]
    \centering
    \caption{Training and validation cases used in this study, with their corresponding Reynolds numbers. Aspect ratios for duct flow cases are denoted as $AR$ and added as subscripts where applicable.}
    \label{tab:training_validation_cases}
    \begin{tabular}{lcc}
        \toprule
        Case & Description & Reynolds Number\\
        \midrule
        \multicolumn{3}{c}{\textbf{Training Cases}} \\
        \midrule
        DF$_{3500, \ AR=1}$  & Duct flow & $\text{Re}_{b} = 3500$ \\
        PH$_{2800}$  & Periodic hill & $\text{Re}_{b} = 2800$ \\
        CF$_{590}$  & Channel flow & $\text{Re}_{\tau} = 590$ \\
        \midrule
        \multicolumn{3}{c}{\textbf{Validation Cases}} \\
        \midrule        
        CF$_{5200}$  & Channel flow & $\text{Re}_{\tau} = 5200$ \\
        DF$_{2600, \ AR=3}$  & Duct flow & $\text{Re}_{b} = 2600$ \\
        CBFS$_{13700}$  & Curved backward-facing step & $\text{Re}_{b} = 13700$ \\
        WMH$_{9.36\cdot10^{5}}$  & Wall-mounted hump & $\text{Re}_{c} = 9.36\cdot10^{5}$ \\
        FP$_{5\cdot10^{6}}$  & Flat-plate boundary layer & $\text{Re}_{b} = 5\cdot10^{6}$ \\
        PH$_{10595}$  & Periodic hills & $\text{Re}_{b} = 10595$ \\
        BUMP$_{20}$  & Bump case & $\text{Re}_{\theta} = 2500$ \\
        BUMP$_{42}$  & Bump case & $\text{Re}_{\theta} = 2500$ \\
        \bottomrule
    \end{tabular}
\end{table*}

To date, there has not been a standardised method for thoroughly validating data-driven turbulence models. However, various studies have utilised a set of CFD cases, such as flat-plate boundary layer development and flow over different obstacles, to assess model performance. In this study, the generalisation capabilities of the trained model are tested across different geometries, Reynolds numbers, and flow characteristics based on prior literature. Hence, the training cases consist of a duct flow case at bulk Reynolds number of 3500 and aspect ratio of 1 (DF$_{3500, \ AR=1}$), flow over periodic hills at bulk Reynolds number of \blue 2800 (PH$_{2800}$)\black, and channel flow at frictional Reynolds number of 590 (CF$_{590}$). Testing cases span from diverse Reynold numbers from the training cases to different geometries as depicted in Table \ref{tab:training_validation_cases}. Notably, the model is tested against the flat-plate FP$_{5\cdot10^{6}}$, and wall-mounted hump WMH$_{9.36\cdot10^{5}}$ \cite{greenblatt2006experimental, greenblatt2006experimental2, naughton2006skin} cases, included in the 2022 NASA Symposium on Turbulence Modelling \cite{rumsey2022nasa}. This initiative challenged the research community to develop and evaluate state-of-the-art data-driven turbulence models, addressing the needs of the field while identifying potential pitfalls and strategies to improve model generalisability and performance.

Each case is analysed by comparing experimental and theoretical data at key locations of velocity, streamwise vorticity, friction coefficient, and relevant Reynolds stress components. 

\subsection{Objective functions}
In order to summarise and simplify the predictions for each case, two different objective functions are defined per case. The first objective ($j_{1}$) defines the improvement of the streamwise velocity. Its definition is case-dependent, following
\begin{subequations}
\begin{equation}
  \text{PH}_{j_{1}} \equiv \text{DF}_{j_{1}} = \frac{\int_{V} \left( | \langle u \rangle - \langle u \rangle^{\text{HF}} | \right) \text{d}V}{ \int_{V} \left( | \langle u \rangle^{k-\omega} - \langle u \rangle^{\text{HF}} | \right) \text{d}V},
        \label{eq:streamwiseObjective1}
\end{equation}    
\begin{equation}
  \text{CF}_{j_{1}} = \frac{\int_{V} \left( | \langle u \rangle - \langle u \rangle^{\text{HF}} | \right) \text{d}V}{ \int_{V} | \langle u \rangle^{k-\omega} | \text{d}V},
        \label{eq:streamwiseObjective2}
\end{equation}
\end{subequations}
where $V$ is the computational volume of each case, $\cdot^{\text{HF}}$ is the field of high-fidelity data, and $\cdot^{k-\omega}$ is the field of standard $k$-$\omega$ SST.

The second objective ($j_{2}$) is a constitutive function to constrain the model's design variables and is specific for each case. These are defined as
\begin{subequations}
\begin{equation}
    \text{PH}_{j_{2}} = \frac{\int_{S} \left( | c_f - c_f^{\text{HF}} | \right) \text{d}S}{\int_{S} \left( | c_f^{k-\omega} - c_f^{\text{HF}} | \right) \text{d}S},
    \label{eq:skin_friction_objective}
\end{equation}    
\begin{equation}
  \text{DF}_{j_{2}} = \frac{\int_{V} \left( | \omega_{1} - \omega_{1}^{\text{HF}} | \right) \text{d}V}{\int_{V} \left( | \omega_{1}^{k-\omega} - \omega_{1}^{\text{HF}} | \right) \text{d}V},
    \label{eq:streamwise_vorticity_objective1}
\end{equation}
\begin{equation}
  \text{CF}_{j_{2}} = \frac{\int_{V} \left( | k - k^{\text{HF}} | \right) \text{d}V}{\int_{V} | k^{k-\omega} | \text{d}V},
    \label{eq:streamwise_vorticity_objective2}
\end{equation}
\end{subequations}

\noindent where $S$ is the bottom wall surface, $c_{f}$ is the friction coefficient, and $\omega_{1}$ is the streamwise vorticity. Based on these definitions, the objective values for all cases except channel flow correspond to 1 when the solution is exact to the standard $k$-$\omega$ SST model, and equal to 0 when the solution is exact to the high-fidelity data. In contrast, for the CF case, the solutions are normalised by the standard $k$-$\omega$ SST prediction since the objective is not to vary its accurate prediction for this case, hence for this case, the values are 0 when the solution is exact to $k$-$\omega$ SST.

\subsection{Training}
To incorporate both PDA corrections into a unified global model, the training process is structured by \textit{progressive} optimisation, which follows a sequential refinement process:

\begin{enumerate}
    \item Optimisation of anisotropy-based secondary flow prediction variables ($A_0$, $A_1$, $A_2$).
    \item Optimisation of separation flow prediction variables ($C_0$, $C_1$, $C_2$, $\lambda_1$, $\lambda_2$).
\end{enumerate}

\begin{table}[t!]
    \centering
    \caption{Weight values $w_i$ assigned to each training case.}
    \label{tab:weights}
    \begin{tabular}{lccc}
        \toprule
         & PH$_{2800}$ & DF$_{3500, \ AR=1}$ & CF$_{590}$ \\
        \midrule
        $w_{i}$ & $0.5$ & $0.5$ & 1 \\
        \bottomrule
    \end{tabular}
    \label{tab:optimisation_weigths}
\end{table}

This sequential approach is adopted to mitigate the inherent numerical instability introduced by the non-linear correction term in the RST anisotropy. The addition of this term can significantly destabilise the flow solver, leading to non-convergent solutions or erroneous velocity field predictions across cases. By initially optimising the secondary flow variables, the model establishes a stable correction baseline for anisotropy-induced phenomena. These variables are then fixed to ensure numerical stability before proceeding to optimise the separation flow variables. This controlled, stepwise refinement enhances robustness and ensures that each correction is systematically integrated without compromising solver stability or convergence.

To integrate all optimisation objectives holistically, a global fitness function is defined as a weighted average to guide the Bayesian optimisation process towards a model with balanced improvements:

\begin{equation}
    J = \frac{1}{N}\sum_{i=1}^{n} j_{i}w_{i},
\end{equation}

\noindent where $J$ represents the global fitness function, $j_{i}$ denotes the specific objective for each case, $w_{i}$ is the corresponding weight, and $N$ is the total number of cases. To ensure equal weighting between secondary flow and separation cases, the assigned weights for each model are shown in Table \ref{tab:optimisation_weigths}.
This weighting strategy ensures a balanced contribution from both secondary flow and separation cases in the optimisation process. Additionally, the value of $J$ is set to 1 when the solution corresponds to the baseline $k$-$\omega$ SST model and 0 when it fully matches high-fidelity reference data.

\blue
A key aspect of this \textit{a posteriori} training methodology is its inherent robustness against overfitting. In this study, overfitting has not been observed in the optimisation runs, which is attributed to the requirement of achieving a converged, steady-state solution for multiple, physically distinct flow scenarios simultaneously. Nonetheless, the need to achieve a converged steady-state solution inherently limits the improvement thresholds that can be achieved, particularly for the non-linear $A_{ij}$ corrections, whose non-linear nature can destabilise the numerical model. Moreover, the inclusion of the channel-flow case (CF$_{590}$) in all optimisation objectives further mitigates overfitting, since any aggressive optimisation towards high-fidelity data in the other cases that would negatively impact the law-of-the-wall prediction is inherently penalised. Hence, each training observation in our methodology incorporates three distinct cases (DF$_{3500, AR=1}$, PH$_{2800}$, and CF$_{590}$), a multi-case strategy which discourages overfitting and promotes a balanced, generalisable model.\black

The progressive optimisation framework enhances model interpretability and generalisability by refining secondary flow and flow separation predictions while maintaining computational efficiency. Importantly, these correction models introduce only a minimal set of algebraic calculations per iteration, ensuring that the computational cost remains comparable to that of the original $k$-$\omega$ SST model.

\section{Results and Discussion}
\label{sec:results}
The results presented in Table \ref{tab:optimisation_results} show that the optimised model exhibits improved global predictions for both velocity ($j_{1}$) and constitutive functions ($j_{2}$). Notably, these enhancements are consistently observed in the validation cases, demonstrating the robustness of the optimised models.
These findings show that the PDA model with a single representative case per correction term is sufficient to enhance model predictions while mitigating the risk of overfitting, thereby preserving both interpretability and generalisability. Furthermore, the results confirm that a simplified progressive optimisation approach yields comparable outcomes to more complex studies as performed in \cite{amarloo2023progressive} while reducing methodological complexity. Given these observations, the subsequent analysis focuses on and shows the detailed results of the optimised model.

\begin{table}[t!]
\centering
\caption{Comparison of the optimised model's objective values across selected cases. Channel flow cases and certain validation cases are excluded, as some validations rely on experimental data for which objective values cannot be directly computed.}
\begin{tabular}{@{}llccc@{}}
\toprule
\multicolumn{1}{c}{\multirow{2}{*}{Usage}} &
  \multicolumn{1}{c}{\multirow{2}{*}{Case}} &
  \multicolumn{3}{c}{$k$-$\omega$ SST PDA} \\ \cmidrule(l){3-5} 
\multicolumn{1}{c}{} & \multicolumn{1}{c}{} & $j_1$  & $j_2$  & $J$             \\ \midrule
Training             & DF$_{3500, \ AR=1}$ & 0.3982 & 0.3554 & 0.3768          \\
Training             & PH$_{2800}$          & 0.3536 & 0.4312 & 0.3924          \\
Validation           & DF$_{2600, \ AR=3}$    & 0.7417 & 0.3720 & 0.5569          \\ 
Validation           & PH$_{10595}$         & 0.3357 & 0.5134 & 0.4246          \\
Validation           & CBFS$_{13700}$       & 0.5776 & 0.3621 & 0.4699          \\
Validation           & BUMP$_{20}$          & 0.9413 & 0.8876 & 0.9145          \\
Validation           & BUMP$_{42}$          & 0.3549 & 0.9317 & 0.6433          \\
\midrule
Average              &                      & 0.5284 & 0.5504 & \textbf{0.5398} \\ \bottomrule
\end{tabular}%
\label{tab:optimisation_results}
\end{table}

\begin{figure}[t!]
    \centering
    \includegraphics[width=1.0\linewidth]{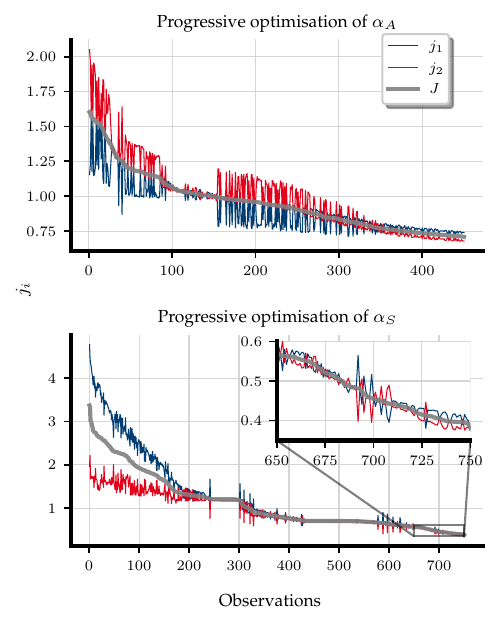}
    \caption{\blue Weighted and global objectives functions of the training cases obtained during the progressive optimisation process.\black}
    \label{fig:Fig4}
\end{figure}

Regarding the training, the adopted Bayesian optimisation strategy shows that consistent minimisation of the objectives is achieved for both corrections. As depicted in Fig. \ref{fig:Fig4}, the model is optimised with 450 observations for the anisotropy correction term $\alpha_{A}$ and 750 observations for both the separation correction term $\alpha_{S}$.

\subsection{Training cases results}
Before evaluating the performance of the developed models in complex flow cases, it is essential to ensure their numerical stability and their ability to reproduce the law-of-the-wall accurately. Without this fundamental validation, further analysis would be unwarranted.

\begin{figure}[t!]
    \centering
    \includegraphics[width=1.0\linewidth]{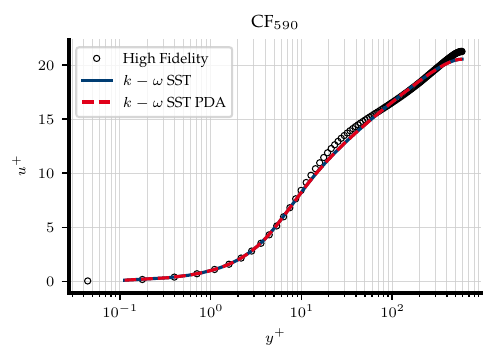}
    \caption{Law-of-the-wall predictions of high-fidelity, standard $k$-$\omega$ SST, and $k$-$\omega$ SST PDA of the CF$_{590}$ training case. High-fidelity data obtained from \cite{lee2015direct}}
    \label{fig:Fig5}
\end{figure}

As shown in Fig.~\ref{fig:Fig5}, the PDA model yields predictions that are identical to those of the standard $k$-$\omega$ SST model, demonstrating that the $\alpha_{A}$ and $\alpha_{S}$ corrections do not compromise the model's baseline accuracy for this case. This consistency arises due to the activation function $\mathcal{X}$, which ensures that the separation correction remains inactive in regions where the ratio $\omega/k$ adheres to the equilibrium assumption. Consequently, the PDA model preserves the original strengths of the $k$-$\omega$ SST formulation while selectively applying corrections only where necessary. Subsequently, the results of the training cases and some of the most relevant validation cases are shown in detail, described, and discussed in order to draw final conclusions.

\begin{figure*}[t!]
    \centering
    \begin{subfigure}[t]{1\textwidth}
        \centering
        \includegraphics[width=1\textwidth]{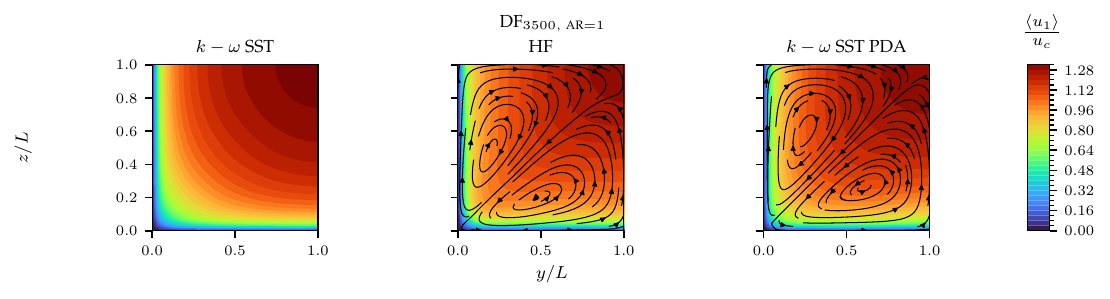}
        \caption{Velocity magnitude contours with stream function.}
        \label{fig:contours_SD_3500_AR1}
    \end{subfigure}
    \hfill \\
    \begin{subfigure}[t]{1\textwidth}
        \centering
        \includegraphics[width=1\textwidth]{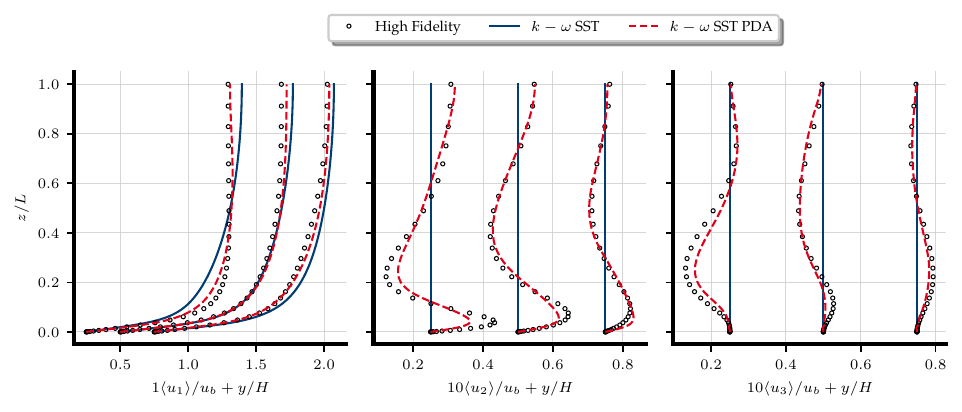}
        \caption{Streamwise, spanwise, and vertical velocity profiles.}
        \label{fig:profiles_SD_3500_AR1}
    \end{subfigure}
    \caption{Qualitative and quantitative analysis of velocity for standard $k$-$\omega$ SST, high-fidelity, and $k$-$\omega$ SST PDA of DF$_{3500, \ AR=1}$ training case. High-fidelity data obtained from \cite{pinelli2010reynolds}.}
    \label{fig:Fig6}
\end{figure*}

Regarding the performance of the model in the square duct flow training case DF$_{3500}$, Fig.\ref{fig:Fig6} illustrates the impact of the PDA $k$-$\omega$ SST model on both velocity distributions due to turbulence anisotropy, comparing results against high-fidelity data and the standard $k$-$\omega$ SST baseline. As expected, the standard $k$-$\omega$ SST model does not predict secondary flow structures, as qualitatively evidenced by the results with the lack of cross-stream streamlines (Fig. \ref{fig:contours_SD_3500_AR1}). In particular, standard $k$-$\omega$ SST neglects the intensity of secondary motions and fails to fully reproduce cross-stream velocity gradients, highlighting its limitations in representing Reynolds stress anisotropy effects. The PDA correction significantly improves these predictions by adjusting the Reynolds stress tensor to account for turbulence-driven secondary flows, reproducing cross-stream velocity gradients and leading to enhanced alignment with high-fidelity data.

Qualitatively, Fig.~\ref{fig:contours_SD_3500_AR1} shows that the PDA-enhanced model improves the prediction of streamwise velocity profiles, especially in the vicinity of the duct walls where secondary flow effects are most pronounced. The baseline model completely neglects any secondary motion, whereas the PDA model corrects this by predicting RST anisotropy more accurately. 
Quantitatively, in Fig.~\ref{fig:profiles_SD_3500_AR1}, the PDA model demonstrates a notable enhancement in predicting the characteristic in-plane vortical structures. The baseline model does not predict a secondary flow field, whereas the PDA correction is capable of doing so, amplifying the in-plane flow effects and more accurately predicting the high-fidelity secondary flow. Importantly, although the separation correction $\alpha_{S}$ is active, the corrections remain inactive for this case, adequately ensuring that improvements are selectively applied to enhance anisotropy representation without introducing spurious modifications elsewhere. \blue It is important to highlight the slight discrepancies between the improved model and the high-fidelity data. Since this is a training case, one could expect that the model's predictions would match the training data, however, due to the pursue of a stable steady-state solution in the CFD solver and the simplification of the full tensor basis expansion for 2D flows (including only $T_{ij}^{(2)}$), the final improvement achieved in the \textit{a posteriori} training performed is constrained to match the high-fidelity data fully.\black

\begin{figure*}[t!]
    \centering
    \begin{subfigure}[t]{1\textwidth}
        \centering
        \includegraphics[width=1\textwidth]{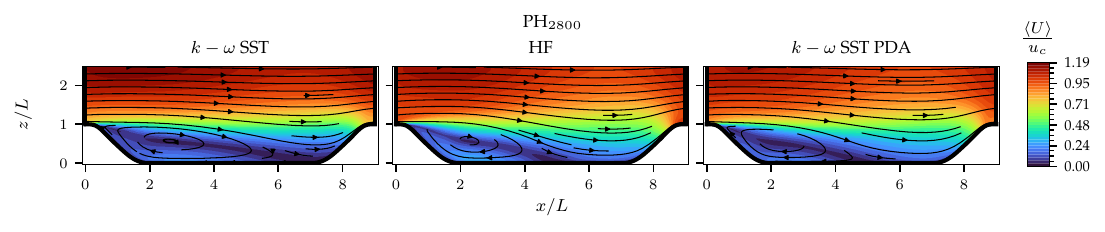}
        \caption{Velocity magnitude contours.}
        \label{fig:Fig7a}
    \end{subfigure}
    \hfill \\
    \begin{subfigure}[t]{1\textwidth}
        \centering
        \includegraphics[width=1\textwidth]{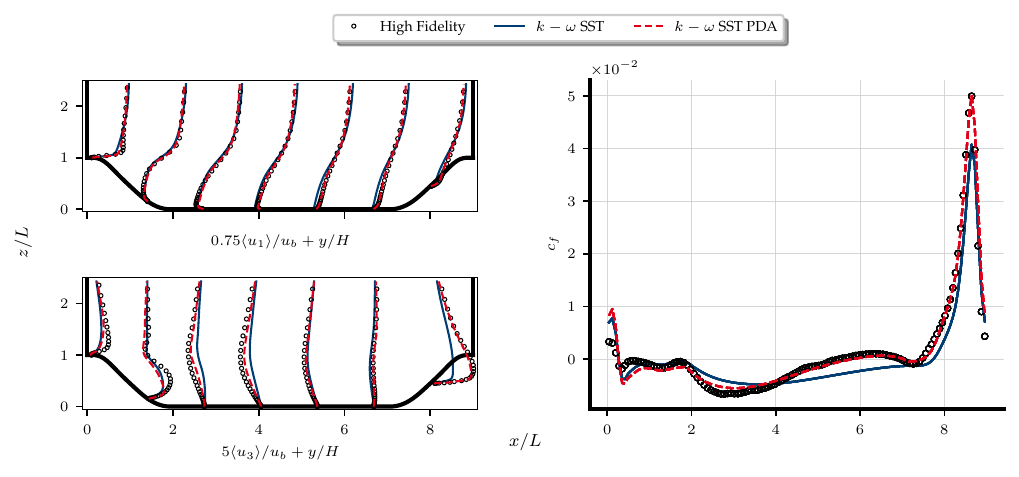}
        \caption{Streamwise and vertical velocity with $c_{f}$.}
        \label{fig:Fig7b}
    \end{subfigure}
    \caption{Qualitative and quantitative analysis of velocity and friction coefficient for standard $k$-$\omega$ SST, high-fidelity, and $k$-$\omega$ SST PDA of PH$_{2800}$ training case. High-fidelity data obtained from \cite{balakumar2015dns}.}
    \label{fig:Fig7}
\end{figure*}

Regarding the performance of the model in the separation training case PH$_{2800}$, the velocity profiles and friction coefficient distributions in Fig.~\ref{fig:Fig7} show the performance of the PDA $k$-$\omega$ SST model against high-fidelity data and the standard $k$-$\omega$ SST baseline. 
The standard $k$-$\omega$ SST model overpredicts the streamwise velocity in the recirculation zone downstream of the hill, reflecting its known tendency to underestimate turbulent viscosity and overextend separation regions.  
The PDA model significantly reduces this discrepancy, aligning closely with the high-fidelity velocity magnitudes in the shear layer and recirculation zone. Improved agreement is particularly evident near reattachment, where the PDA correction adjusts $\nu_t$ to improve the prediction of flow recovery dynamics.
Furthermore, the baseline model underestimates the magnitude of $c_f$ in the separation region, failing to resolve the sharp recovery of wall shear stress post-reattachment and on top of the hill observed in high-fidelity data.  
The PDA-enhanced model corrects this behaviour, reproducing the $c_f$ trough depth and subsequent rise with enhanced fidelity. The optimised separation correction locally increases $\nu_t$, reducing the overpredicted recirculation length and improving shear stress gradients.
In the attached flow regions upstream of separation, the PDA model retains near-identical predictions to the standard $k$-$\omega$ SST, confirming that corrections remain inactive where the baseline performs adequately.

\begin{figure*}[t!]
    \centering
    \includegraphics[width=1\linewidth]{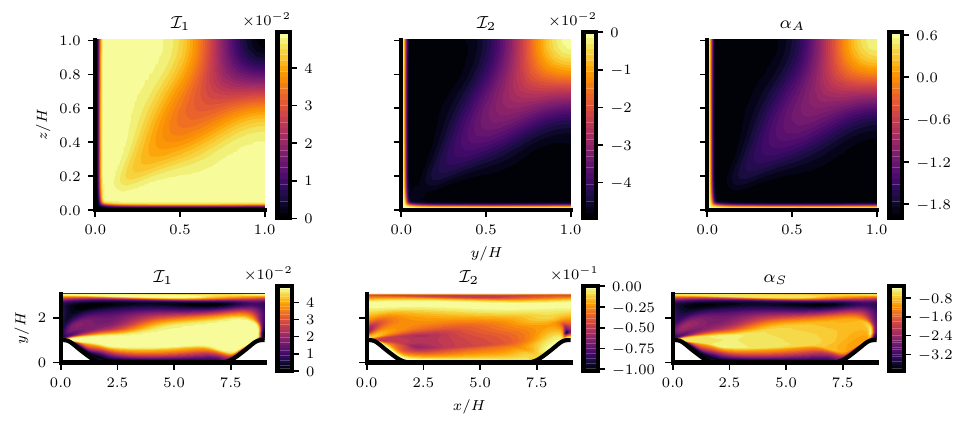}
    \caption{\blue Invariant and correction functions values for the two training cases DF$_{3500, \ AR=1}$ (top row) and PH$_{2800}$ (bottom row).\black}
    \label{fig:Fig8}
\end{figure*}

\begin{table*}[t!]
    \centering
    \caption{Coefficient values for the developed model.}
    \label{tab:model_coefficients}
    \begin{tabular}{llllllll}
        \toprule
        $A_{0}$ & $A_{1}$ & $A_{2}$ & $C_{0}$ & $C_{1}$ & $C_{2}$ & $\lambda_{1}$ & $\lambda_{2}$ \\
        \midrule
        -1.584 & -0.685 & -0.178 & -2.070 & 1.119 & -0.215 & 18.622 & 4.698 \\
        \bottomrule
    \end{tabular}
\end{table*}

\blue It is noted that a discrepancy in skin friction persists in the attached flow region on the crest of the hill, where both the baseline and the PDA model over-predict the high-fidelity data. This behaviour is a known characteristic of RANS models applied to this periodic flow case. The PDA model is designed as a targeted correction for separated flows, and its activation function, $\mathcal{X}$, is explicitly intended to confine corrections to regions where the baseline model struggles most. The observation that the PDA model's prediction mirrors the baseline $k-\omega$ SST in this attached region is therefore not a failure, but rather a confirmation that the activation function is operating as intended. Furthermore, the mesh resolution used for the \textit{a posteriori} RANS optimisation is necessarily far coarser than that of the reference DNS \cite{balakumar2015dns}; achieving local DNS-level accuracy on a RANS grid is computationally prohibitive and not the objective of this framework. The primary goal (improving the prediction of the downstream separation) is successfully achieved, demonstrating a robust and practical advancement.\black

\blue
To better illustrate the spatial impact and interpretable nature of the PDA model's corrections, Fig. \ref{fig:Fig8} displays contours of the key correction components for both the DF${3500, AR=1}$ and PH${2800}$ cases and Table \ref{tab:model_coefficients} shows the final coefficient values optimised for the model. For the duct flow, it is observed that the flow invariants $\mathcal{I}_1$ and $\mathcal{I}_2$ exhibit distinct anticorrelated spatial patterns in the corner regions. When combined to form the anisotropy correction $\alpha_A$, the resulting spatial structure strongly resembles that of the second invariant, $\mathcal{I}_2$, highlighting its importance in capturing the features of the secondary flow. For the periodic hill case, which drives the separation correction, the invariants are concentrated in the shear layer detaching from the hill crest. Here, the structure of the resulting separation correction, $\alpha_S$, is clearly dominated by the contribution from $\mathcal{I}_1$, which is consistent with the larger magnitude of its corresponding coefficient, $C_1$, in Table \ref{tab:model_coefficients}. This analysis visually demonstrates how the framework uses physically-meaningful, spatially-varying inputs to build targeted corrections for specific flow phenomena.

The mathematical structure of the corrections $\alpha_A$ and $\alpha_S$, as linear combinations of invariants plus a constant term, is explicit and transparent. The inputs to these corrections, specifically the invariants $I_1, I_2$, are derived from well-defined, physically meaningful local flow quantities ($S_{ij}, \Omega_{ij}, \omega$); while the activation function $\mathcal{X}$ is based on quantities ($k, \omega, \nu_t$) that characterise the state of turbulence and its deviation from equilibrium. The coefficients ($A_i, C_i, \lambda_i$), while determined through data-driven optimisation, act as fixed weights for these physically interpretable terms. The final values of the coefficients $A_i$ and $C_i$ (presented in Table \ref{tab:model_coefficients}) provide an indication of the learned importance and direction of influence of the respective standardised invariants on the corrections $\alpha_A$ and $\alpha_S$. This approach is distinct from less transparent models (where the entire correction might be, for example, the direct output of a complex neural network). 

\black

\subsection{Validation data}
\begin{figure}[t!]
    \centering
    \includegraphics[width=1.0\linewidth]{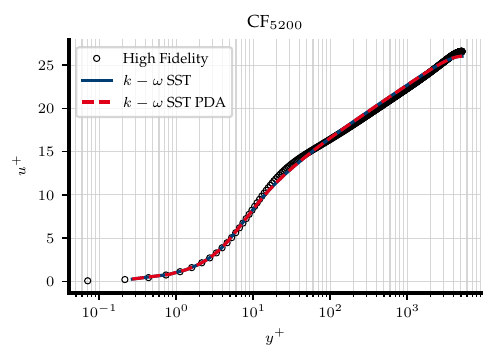}
    \caption{Law-of-the-wall velocity for high-fidelity, standard $k$-$\omega$ SST, and $k$-$\omega$ SST PDA of the CF$_{5200}$ validation case. High-fidelity data obtained from \cite{lee2015direct}}
    \label{fig:Fig9}
\end{figure}
As the model has been extensively validated across multiple test cases, only the most representative results are presented in the main sections of this study. For a more general overview of the additional results, the reader is referred to Table \ref{tab:optimisation_results}, where the final values for each objective are shown.

\begin{figure*}[t!]
    \centering
    \begin{subfigure}[t]{1\textwidth}
        \centering
        \includegraphics[width=1\textwidth]{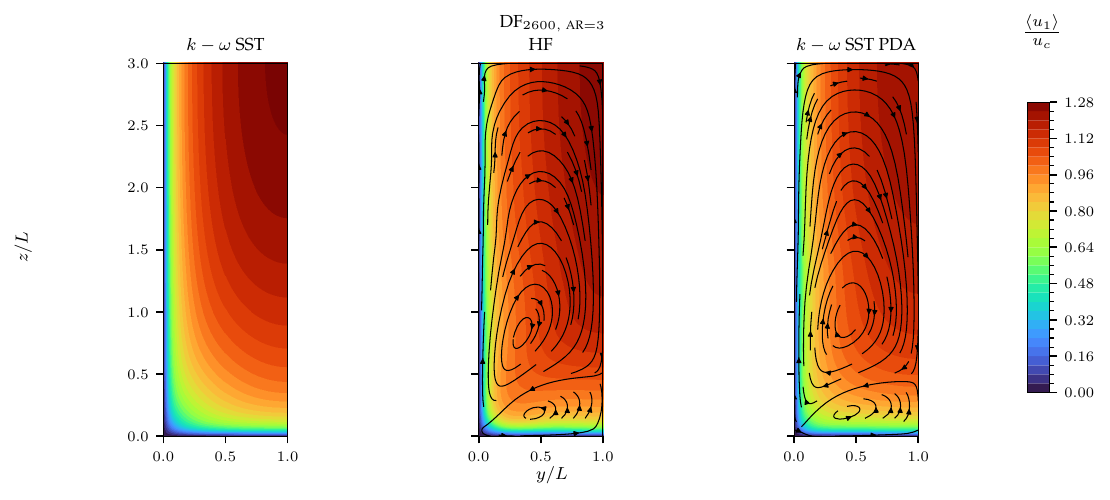}
        \caption{Velocity magnitude contours with stream function.}
        \label{fig:Fig10a}
    \end{subfigure}
    \hfill \\
    \begin{subfigure}[t]{1\textwidth}
        \centering
        \includegraphics[width=1\textwidth]{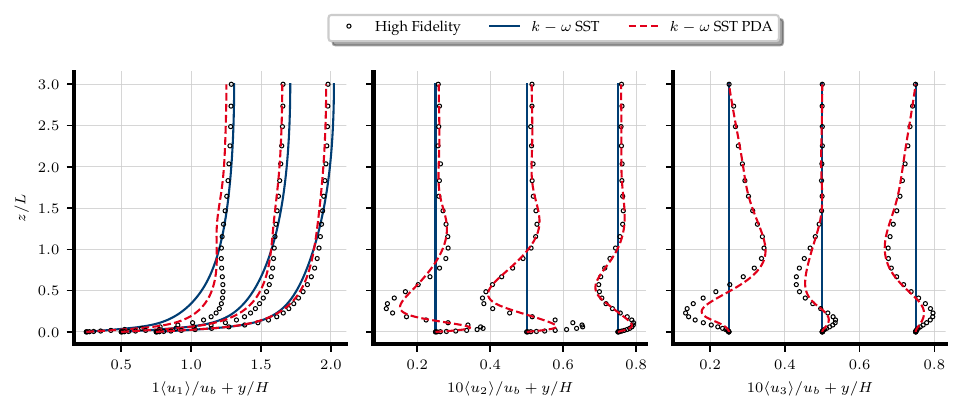}
        \caption{Streamwise, spanwise, and vertical velocity profiles.}
        \label{fig:Fig10b}
    \end{subfigure}
    \caption{Qualitative and quantitative analysis of velocity for standard $k$-$\omega$ SST, high-fidelity, and $k$-$\omega$ SST PDA of DF$_{2600, \ AR=3}$ validation case. High-fidelity data obtained from \cite{pinelli2010reynolds}.}
    \label{fig:Fig10}
\end{figure*}

To first ensure the validity of this study, it is essential to verify that the model accurately predicts the law-of-the-wall and generalises to cases where the standard $k$-$\omega$ SST model provides reliable results. Therefore, an initial comparison is made against a channel flow case with a higher Reynolds number (CF$_{5200}$), as shown in Fig.~\ref{fig:Fig9}. The results show an exact agreement with the predictions of the standard $k$-$\omega$ SST model, confirming that the PDA corrections do not introduce unintended modifications in well-predicted flow cases. 

Hence, the performance and generalisability of the Reynolds stress anisotropy corrections are further assessed using a duct flow case with a different Reynolds number and a higher aspect ratio, DF$_{2600, \ AR=3}$. Consistent with the trends observed in the training case DF$_{3500, \ AR=1}$, the PDA model significantly improves predictions compared to the standard $k$-$\omega$ SST formulation, as depicted in Fig. \ref{fig:Fig10}. Both qualitatively and quantitatively, the secondary motions are accurately predicted, including their correct directional behaviour. However, in regions with high-velocity gradients near the walls, slight discrepancies in magnitude persist when compared to high-fidelity data. Nevertheless, a substantial overall improvement is achieved while maintaining numerical stability and computational robustness.

To further assess the generalisability and the separation correction $\alpha_{S}$, five validation cases are considered: a higher Reynolds number case PH$_{10595}$ and four additional cases --- CBFS$_{13700}$, BUMP$_{20}$, BUMP$_{42}$, and NASA challenge case WMH$_{9.36\cdot10^5}$. For conciseness, only the detailed results for CBFS$_{13700}$ and WMH$_{9.36\cdot10^5}$ are presented here, while final results for PH$_{10595}$, BUMP$_{20}$, and BUMP$_{42}$ are provided in Table \ref{tab:optimisation_results} \blue and \ref{app:A}\black.

\begin{figure*}[t!]
    \centering
    \begin{subfigure}[t]{1\textwidth}
        \centering
        \includegraphics[width=1\textwidth]{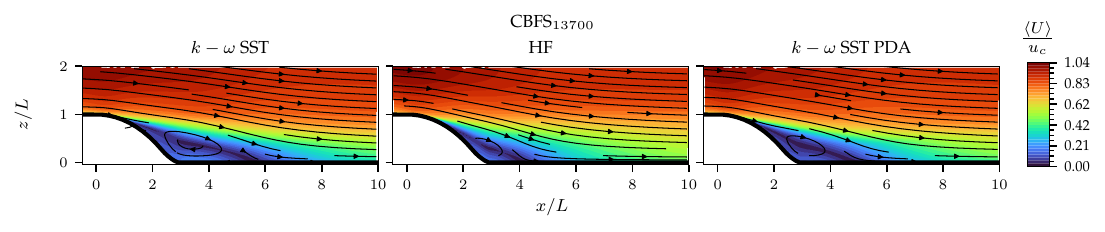}
        \caption{Velocity magnitude contours.}
        \label{fig:Fig11a}
    \end{subfigure}
    \hfill \\
    \begin{subfigure}[t]{1\textwidth}
        \centering
        \includegraphics[width=1\textwidth]{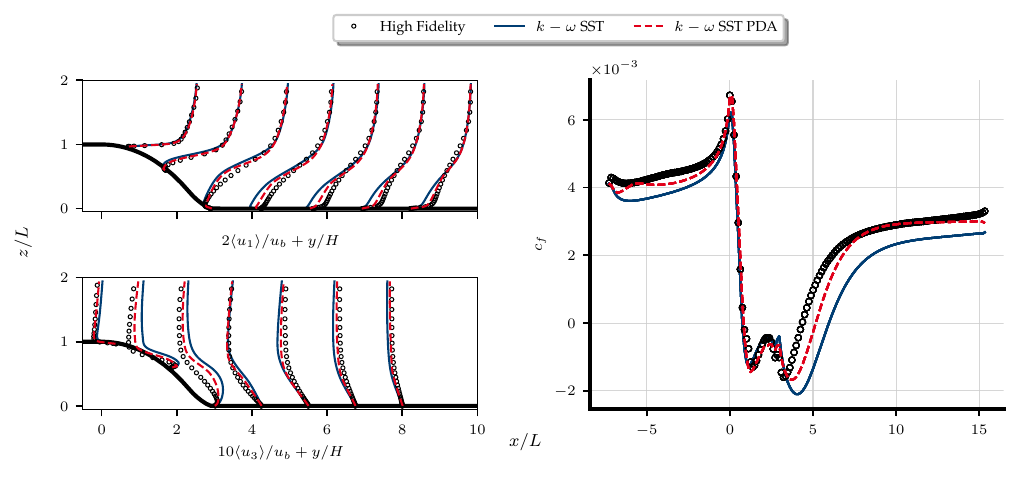}
        \caption{Streamwise and vertical velocity with $c_{f}$.}
        \label{fig:Fig11b}
    \end{subfigure}
    \caption{Qualitative and quantitative analysis of velocity and friction coefficient for standard $k$-$\omega$ SST, high-fidelity, and $k$-$\omega$ SST PDA of CBFS$_{13700}$ validation case. High-fidelity data obtained from \cite{bentaleb2012large}.}
    \label{fig:Fig11}
\end{figure*}

The performance of the PDA $k$-$\omega$ SST model in all separation validation cases closely follows the trends observed in the training case PH$_{2800}$. As shown in Fig.~\ref{fig:Fig11}, velocity profiles and friction coefficient distributions for CBFS$_{13700}$ exhibit the same enhanced predictive capability achieved in the training cases, confirming the robustness of the correction in accurately capturing separation-induced flow features. Similar to PH$_{2800}$, the standard $k$-$\omega$ SST model overpredicts the streamwise velocity in the recirculation region, reflecting its known tendency to underestimate turbulent viscosity and extend separation zones beyond the observed high-fidelity data. The PDA correction mitigates this discrepancy by refining the turbulence transport mechanisms, leading to improved agreement in the shear layer and reattachment region.

\blue To ensure a direct and fair comparison, the simulations were initialised by interpolating the velocity profiles from the high-fidelity reference data directly at the inlet plane. This is why the skin-friction values for all models match the high-fidelity data precisely at the start of the domain, as seen in Fig. \ref{fig:Fig11b}. However, a RANS model has its own internal equilibrium for a turbulent boundary layer, which naturally differs from that of the DNS. Hence, immediately downstream of the inlet, the SST model adjusts the flow from the imposed state towards its own equilibrium, creating a difference in results for $c_f$ comparwed to high-fidelity data. This method was chosen deliberately to guarantee that all simluations for this case start from the exact same conditions, thereby isolating the performance differences purely to the turbulence models themselves. \black

\blue The wall-mounted hump case (WMH$_{9.36\cdot10^5}$) is included as a particularly challenging test of the model's extrapolative capabilities. This case is a well-established benchmark within the turbulence modelling community, partly due to documented difficulties in reproducing the exact experimental conditions, particularly in the upstream attached flow, where simulations often mismatch experimental skin friction \cite{rumsey2022nasa}. The PDA model's separation correction is designed primarily to improve the local physics of the separation bubble and reattachment phenomena, most directly reflected in the skin friction ($c_f$) distribution. The pressure coefficient ($c_p$), being a more globally-influenced quantity, is expected to be less sensitive to such localised corrections.

\begin{figure}[t!]
    \centering
    \begin{subfigure}[t]{1\linewidth}
        \centering
        \includegraphics[width=1\textwidth]{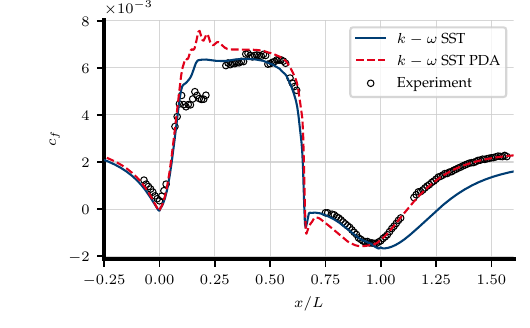}
        \caption{Friction coefficient.}
        \label{fig:Fig12a}
    \end{subfigure}
    \begin{subfigure}[t]{1\linewidth}
        \centering
        \includegraphics[width=1\textwidth]{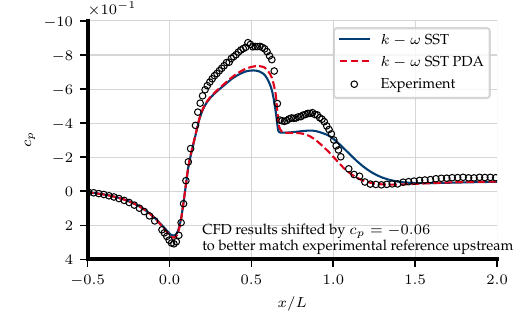}
        \caption{Pressure coefficient.}
        \label{fig:Fig12b}
    \end{subfigure}
    \caption{Results for testing case WMH$_{9.36\cdot10^5}$ following \cite{rumsey2022nasa}. Reference experimental data from \cite{greenblatt2006experimental, greenblatt2006experimental2, naughton2006skin}.}
    \label{fig:Fig12}
\end{figure}

The results for this case are presented in Fig.~\ref{fig:Fig12}. Despite the case's inherent complexities and the noted upstream mismatch, the PDA model shows a consistent improvement where the correction is designed to be active. The baseline model's underestimation of the $c_f$ magnitude in the separation region and its failure to capture wall-shear stress recovery are significantly improved by the PDA correction. This is achieved by locally increasing $\nu_t$, which reduces the overprediction of the recirculation length. The recovery of $c_f$ following separation closely matches the experimental data, demonstrating that the local correction is working as intended. For the pressure coefficient ($c_P$) (Fig. \ref{fig:Fig11b}), the results remain largely consistent with the predictions of the standard $k$-$\omega$ SST model, though slight improvements are observed at $x/L > 0.6$, where the PDA-enhanced predictions align more closely with the experimental data. Evaluating the model against this complex case thus provides a valuable demonstration of the correction's targeted effectiveness and robustness.\black

Across all validation cases, the PDA model consistently enhances predictions without introducing unintended modifications in regions where the baseline model already performs adequately. This outcome further demonstrates the reliability of the separation correction $\alpha_{S}$ in improving separation-dominated flows while maintaining numerical stability and preserving the underlying strengths of the $k$-$\omega$ SST formulation.

\begin{figure}[t!]
    \centering
    \begin{subfigure}[t]{1\linewidth}
        \centering
        \includegraphics[width=1\textwidth]{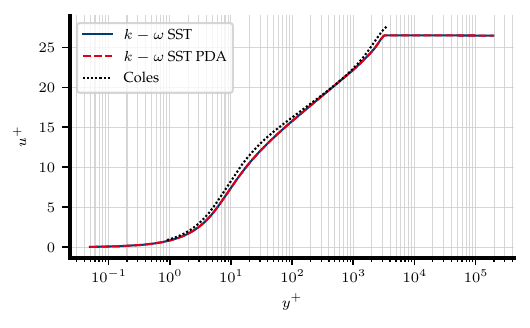}
        \caption{Boundary layer on the finest 513 $x$-grid point for FP$_{5\cdot10^{6}}$.}
        \label{fig:Fig13a}
    \end{subfigure}
    \begin{subfigure}[t]{1\linewidth}
        \centering
        \includegraphics[width=1\textwidth]{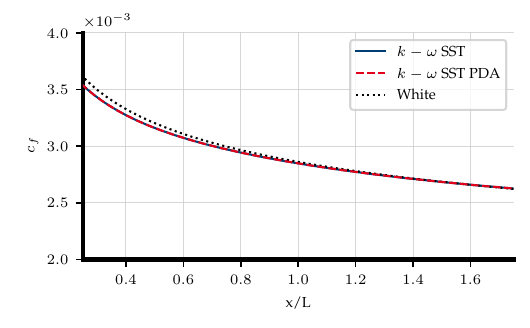}
        \caption{Friction coefficient and normalised streamwise distance for FP$_{5\cdot10^{6}}$.}
        \label{fig:Fig13b}        
    \end{subfigure}
    \caption{\blue Simplified results for testing case FP$_{5\cdot10^6}$.\black}
    \label{fig:Fig13}
\end{figure}

To establish a final validation assessment, the model is further examined where the anisotropy and separation corrections are not expected to influence the flow predictions. In these cases, the specific physical effects that the model's corrections are designed to address are absent. Consequently, the PDA-enhanced formulation should yield results that are identical to the standard $k$-$\omega$ SST model. Any significant deviation would indicate unintended modifications to well-predicted flow regions. In order to assess this influence, the FP$_{5\cdot10^{6}}$ case is evaluated. As shown in Fig.~\ref{fig:Fig13}, the PDA model produces predictions that are indistinguishable from those of the standard $k$-$\omega$ SST model for both the streamwise velocity and the friction coefficient. \blue To compare the friction coefficient, White's \cite{white2006viscous} correlation $c_f=0.455\left( \ln{(0.06\text{Re}_{x})} \right)^{-2}$ is used. This correlation has been studied by Nagib et al. (2007) \cite{nagib2007approach} and showed great agreement between experimental and high-fidelity numerical data\black. This outcome further demonstrates the robustness of the activation function $\mathcal{X}$, which effectively prevents corrections from being applied in regions where they are unnecessary, ensuring that the baseline model's accuracy is preserved.

These results validate the PDA framework's ability to enhance separation and anisotropy-based secondary flow predictions while preserving the robustness of the original model, addressing a critical limitation of the $k$-$\omega$ SST formulation for adverse pressure gradient flows and non-linear Reynolds stress corrections.

\section{Conclusions}
\label{sec:conclusions}
This study presents a systematic framework for enhancing the interpretability and generalisability of the standard $k$-$\omega$ SST turbulence model through progressive data augmentation and \textit{a posteriori} CFD-driven Bayesian optimisation. By integrating non-linear Reynolds stress anisotropy corrections and activation-based separation corrections within a multi-case optimisation strategy, the developed model achieves improved predictive accuracy across diverse two-dimensional flow regimes while preserving the numerical stability and robustness of the baseline formulation. The progressive augmentation methodology ensures that the original predictive capabilities of the $k$-$\omega$ SST model in canonical flows—such as channel flow and attached boundary layers are rigorously maintained. The corrections remain inactive in regions where the baseline model performs adequately, thereby avoiding detrimental interference with its validated strengths.

The introduction of a data-driven RST anisotropy correction term, derived from Pope's decomposition \cite{pope1975more}, significantly improves the prediction of secondary flows in duct geometries. This correction addresses the limitations of linear eddy-viscosity assumptions without introducing unphysical artefacts or compromising numerical stability. Furthermore, an activation function, optimised via Bayesian methods, enables localised implicit corrections to the turbulent eddy viscosity $\nu_{t}$ by modifying the $\omega$-equation under adverse pressure gradient conditions. This mechanism robustly increases turbulent viscosity in separation-prone regions, reducing overprediction of recirculation zones while retaining equilibrium behaviour in attached boundary layers.

Generalisability across diverse Reynolds numbers and validation against unseen cases, including high Reynolds number boundary layers, flow over diverse obstacles, and flat-plate boundary-layer development, demonstrate consistent improvements in velocity field, streamwise vorticity, and skin friction predictions. The model adapts effectively to varying geometries and flow conditions, underscoring its capacity to generalise and extrapolate beyond training datasets. Moreover, the use of sparse regression and physics-informed candidate functions based on the first two flow invariants ensures that corrections remain consistent and aligned with turbulence scaling laws. This approach enhances model transparency, facilitating direct analysis of correction mechanisms and their contributions to flow physics.

By embedding corrections within algebraic expressions, the augmented model incurs negligible computational overhead compared to the standard $k$-$\omega$ SST formulation. This efficiency ensures practical applicability to industrial-scale simulations. These advancements provide a promising solution to data-driven turbulence modelling, particularly the trade-off between accuracy and generalisability. The methodology's success in balancing data-driven enhancements with physical constraints highlights the potential of progressive augmentation as a paradigm for developing robust, interpretable, and deployable RANS models. Future work could extend this framework to three-dimensional flows and incorporate model discovery to enhance the chosen candidate functions of the model. By bridging the gap between machine learning techniques and traditional turbulence modelling, this study contributes a validated pathway toward reliable, practically viable CFD tools.

\section*{CRediT authorship contribution statement}
\textbf{M. J. Rinc\'on}: Software, Investigation, Validation, Formal analysis, Visualisation, Writing – original draft. \textbf{M. Reclari}: Conceptualisation, Methodology, Project administration, Resources, Supervision, Writing – review \& editing. \textbf{X. I. A. Yang}: Conceptualisation, Methodology, Supervision, Writing – review \& editing. \textbf{M. Abkar}: Conceptualisation, Methodology, Project administration, Resources, Supervision, Writing – review \& editing.

\section*{Declaration of Competing Interest}
The authors declare that they have no known competing financial interests or personal relationships that could have appeared to influence the work reported in this paper.

\section*{Acknowledgments}
This research is supported by Innovation Fund Denmark (IFD) under Grant No. 3130-00007B and Kamstrup A/S. 
X. I. A. Yang acknowledges Air Force Office of Scientific Research award number FA9550-23-1-0272. 
The authors would also like to acknowledge the EuroHPC Joint Undertaking for awarding this project access to the EuroHPC supercomputer LUMI, hosted by CSC (Finland) and the LUMI consortium through a EuroHPC Regular Access call. This work was also partially supported by the Danish e-Infrastructure Cooperation (DeiC) National HPC under grant number DeiC-AU-N5-2025130.

\section*{Data availability}
The CFD model developed for this study is publicly available for OpenFOAM at \href{https://github.com/AUfluids/KOSSTPDA.git}{Aarhus University Fluids GitHub repository} and as part of the OpenFOAM Turbulence Technical Committee Repository. Further data will be made available on request.


\appendix
\blue
\section{Additional validation results}
\label{app:A}
In this section, the three validation cases that have not been discussed in the main text are shown. These results have not been shown in the main text due to their similarity with previous results and not to overcrowd the paper with figures. Otherwise, the main thread of the study can be challenging to follow.

\begin{figure*}[t!]
    \centering
    \includegraphics[width=1\linewidth]{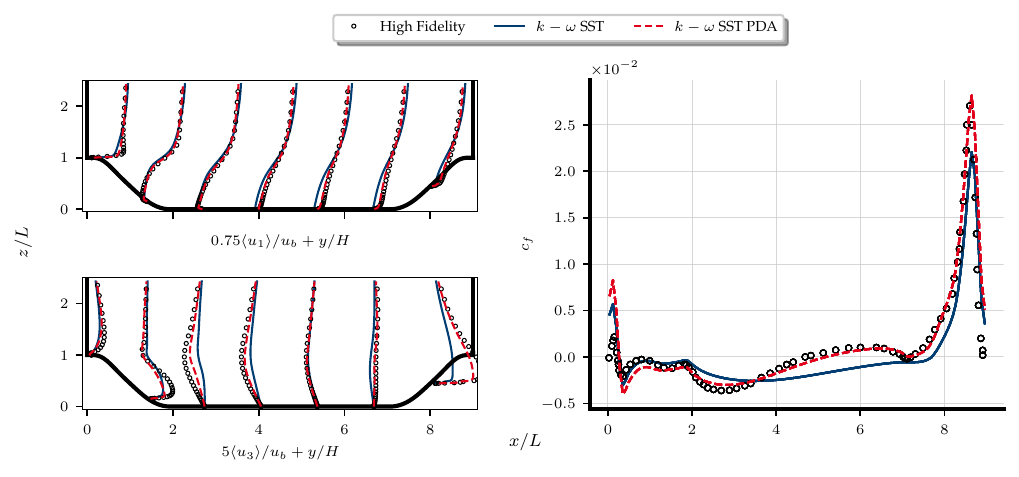}
    \caption{Quantitative analysis of velocity and friction coefficient for standard $k$-$\omega$ SST, high-fidelity, and $k$-$\omega$ SST PDA of PH$_{10595}$ testing case. High-fidelity data obtained from \cite{balakumar2015dns}.}
    \label{fig:FigA1}
\end{figure*}

\begin{figure*}[t!]
    \centering
    \begin{subfigure}[t]{1\textwidth}
        \centering
        \includegraphics[width=1\textwidth]{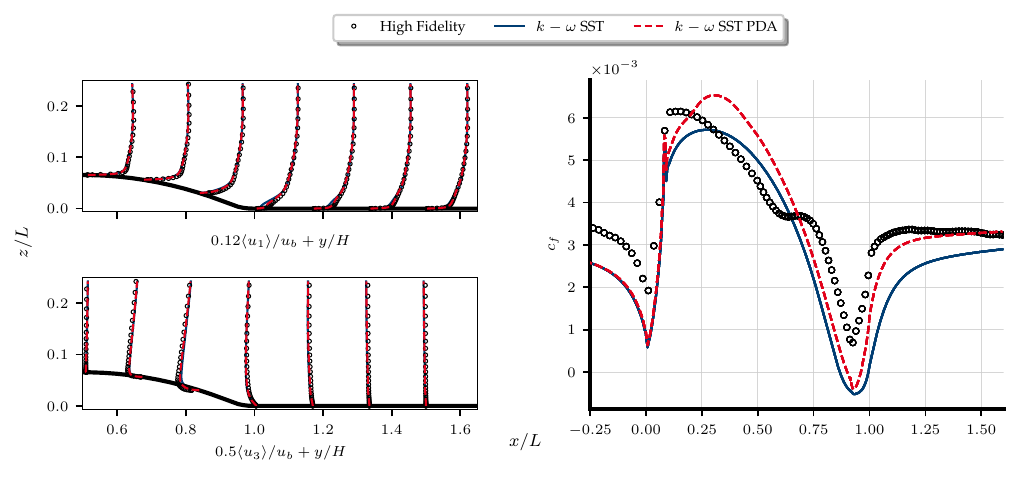}
        \caption{BUMP$_{20}$.}
        \label{fig:FigA2a}
    \end{subfigure}
    \hfill \\
    \begin{subfigure}[t]{1\textwidth}
        \centering
        \includegraphics[width=1\textwidth]{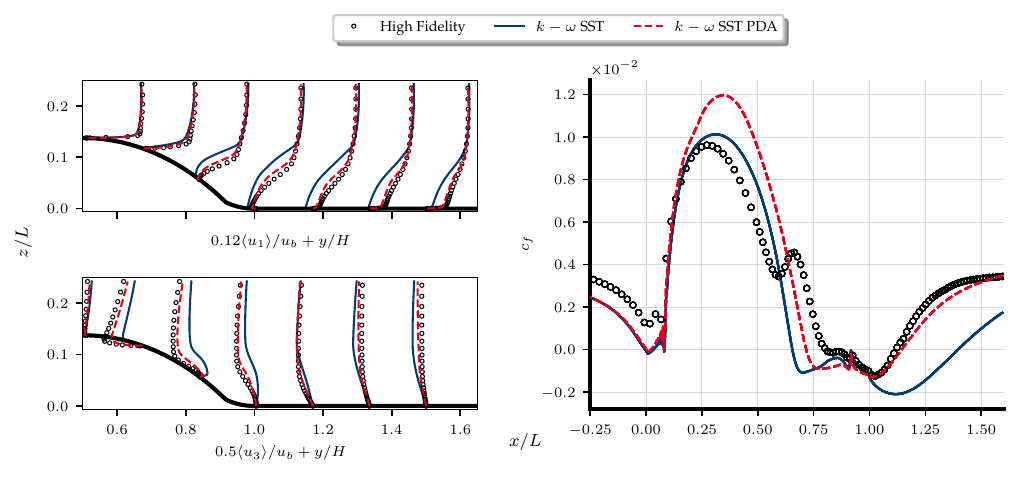}
        \caption{BUMP$_{42}$}
        \label{fig:FigA2b}
    \end{subfigure}
    \caption{Quantitative analysis of velocity and friction coefficient for standard $k$-$\omega$ SST, high-fidelity, and $k$-$\omega$ SST PDA of BUMP$_{20}$ and BUMP$_{42}$ testing cases. High-fidelity data obtained from \cite{matai2019large}.}
    \label{fig:FigA2}
\end{figure*}

Results for the case PH$_{10595}$ are shown in Fig. \ref{fig:FigA1}, whereas results for cases BUMP$_{20}$ and BUMP$_{42}$ are shown in Figs. \ref{fig:FigA2a} and \ref{fig:FigA2b} respectivelity. For all cases, an overall localised improvement is seen from the PDA model compared to the standard $k$-$\omega$ SST model. Even in cases where no flow separation is predicted by high-fidelity data (case BUMP$_{20}$), the friction coefficient recovery downstream of the adverse pressure gradient region, is predicted with higher accuracy.

\black

%\bibliographystyle{cas-model2-names}
\bibliographystyle{elsarticle-num} 

\bibliography{cas-refs}

\end{document}